# Validity conditions of direct boundary integral equation for exterior problems of plane elasticity

Alain Corfdir[a], Guy Bonnet[b]

[a]CERMES, Institut Navier, ENPC, 6 et 8 avenue Blaise Pascal, 77455 Marne la Vallée, France

[b]Université de Marne la Vallée, Laboratoire de Mécanique, Institut Navier, 5 boulevard Descartes, 77454 Marne la Vallée cedex, France

**Abstract** Writing the boundary integral equation for an exterior problem of elasticity is subordinate so far to hypotheses on the asymptotical behaviour at infinity of solutions. The sufficient conditions met in the literature are too restrictive and do not notably cover the case when the loading has a non zero resultant force. This difficulty can be removed by considering the problem in displacements relatively to one point located at a finite distance from the loading. Finally, this auxiliary problem allows widening the conditions of validity of the usual formulation of the direct integral method.

**Résumé Conditions de validité de l'équation intégrale directe pour les problèmes extérieurs de l'élasticité plane** L'établissement de l'équation intégrale frontière pour un problème extérieur d'élasticité nécessite des hypothèses sur le comportement à l'infini des solutions en déplacements et en contraintes. Les conditions suffisantes établies jusqu'ici sont trop restrictives et ne couvrent pas le cas d'un chargement ayant une résultante non nulle. Cette difficulté est écartée en considérant un problème en déplacement relatif. Enfin, ce problème auxiliaire permet d'étendre les conditions de validité de la formulation usuelle de la méthode intégrale directe.

Keywords: Statics; Plane elasticity; Boundary methods; Direct method

# 1. Introduction

Engineering applications justify considering the exterior problem of a half-space or a half-plane in elasticity. The extension of civil works is indeed small compared to the one of the soil mass, which can be considered as infinite and 2D analyses are of frequent use in geotechnical engineering. One faces however the difficulty of the asymptotic behaviour of the fundamental solution for plane elasticity which increases as a logarithm at infinity. The solution of plane elasticity problems on non-bounded domains requires specific conditions at infinity and appears paradoxical (e.g., [1]). One also has to tackle the difficulty of writing valid boundary integral equations and integral representations. The purpose of the paper is to justify the use of the boundary integral equation method for any case of loading at the boundary, including the case of loading having a non zero resultant.

# 2. Sufficient conditions for obtaining a boundary integral equation related to 2D elasticity problems within previous works

Let us consider an open part $\overline{\Omega}$ of a linear elastic, isotropic, homogeneous half-plane, a bounded part $\Omega$ of which having been removed (Fig. 1a). There are no volume forces. To apply the direct method, the boundary integral equation is written on the part of the bounded boundary $\partial\overline{\Omega}$ where displacements or tractions are prescribed. Such an integral equation is obtained by

- writing the integral equation on the domain comprised between $\partial\overline{\Omega}$ and a half-circle $S_r$
- looking for the limit of the integrals on $S_r$ when its radius r tends to infinity.

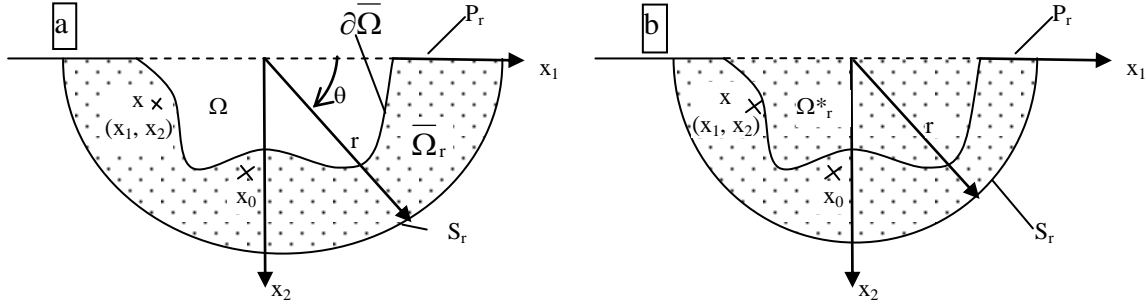

**Figure 1 a: Integration domain $\overline{\Omega}_R$ (in grey) b: auxiliary integration domain $\Omega^*_r$**

Figure 1 a : Domaine d'intégration $\overline{\Omega}_R$ (en grisé) b : domaine auxiliaire d'intégration $\Omega^*_r$

Under assumptions on the behaviour of the solution at infinity, a regular enough solution satisfies a boundary integral equation (1) for a regular ($C^2$) point of the boundary, and an integral representation (2) [e.g., 2]:

$$u_k(x) + \int_{\partial\overline{\Omega}} \{(u_i(y) - u_i(x))T_i^k(x,y) - t_i(y)U_i^k(x,y)\}dS_y = 0 \quad x \in \partial\overline{\Omega} \tag{1}$$

$$u_k(x) = \int_{\partial\overline{\Omega}} \{t_i(y)U_i^k(x,y) - u_i(y)T_i^k(x,y)\}dS_y \quad x \in \overline{\Omega} \tag{2}$$

The solutions in displacement and traction are denoted by u and t. Functions U and T are elementary solutions in displacement and traction for the half plane. The functions $U_i^k$ are defined up to an arbitrary translation. The usual choice [e.g. 2], which is adopted here, corresponds to behaviour at infinity such that:

$$U_i^2((x_1, x_2 = 0), y) = A_i^2 \ln|x_1| + B_i^2 \text{sign}(x_1) + O(1/x_1) \quad x_1 \to \pm\infty \tag{3}$$

$$U_i^1((x_1 = 0, x_2), y) = A_i^1 \ln(x_2) + O(1/x_2) \quad x_2 \to +\infty \tag{4}$$

Where $A_i^j$ and $B_i^j$ are constants.

In the 2D case, different authors proposed sufficient conditions on the behaviour of the solution at infinity so that it satisfies a boundary integral equation on $\partial\overline{\Omega}$. Watson [3] gave: u(x)=o($r^{-1}$) and $\sigma = o(r^{-2})$; Maiti and al. [4] u(x)=O($r^{-1}$) $\sigma = O(r^{-2})$. Constanda [5, 6] and Schiavone and Ru [7] used the hypothesis that $u_i$ decrease at infinity as $r^{-1}(a\cos\theta + b\sin\theta + c\cos(3\theta) + d\sin(3\theta)) + O(r^{-2})$. Bonnet [2] gave a less restrictive sufficient condition: $u(x) = O(r^{-\alpha})$ and $\sigma = O(r^{-1-\alpha})$ with $\alpha > 0$.

In conclusion, it seems that the least restrictive sufficient condition that is presently known is the one given by [2]. All the sufficient conditions described above are not at all satisfying because they cannot justify studying the boundary problem related to a point loading or (principle of Saint-Venant) a loading with a non zero resultant force. The purpose of the following is to show that the classical boundary integral equations (1, 2) are also valid if the resultant of applied forces is non-zero.

## 3. Integral equation and integral representation related to the relative displacement

Poulos and Davis [8] stated that: "displacements due to line loading on or in a semi-infinite mass are only meaningful if evaluated as the displacement of one point relatively to another point, both points being located neither at the origin of loading nor at infinity". Accordingly, to mitigate the difficulties related to the behaviour at infinity, it seems natural to introduce the relative displacements in the formulation of the problem. To this aim, a first step is to build a boundary integral equation whose solution corresponds to displacements with respect to a reference point $x_0$ taken within $\overline{\Omega}$ (outside the boundary $\partial\overline{\Omega}$). This is equivalent to setting a supplementary condition of no displacement for this reference point $x_0$.

The boundary conditions correspond to prescribed displacements on $\partial\overline{\Omega}_U$ and prescribed tractions on $\partial\overline{\Omega}_F$, $\partial\overline{\Omega}_U$ and $\partial\overline{\Omega}_F$ being complementary parts of $\partial\overline{\Omega}$. The displacement is zero at point $x_0$. Hence the following conditions are met:

$$u(x) = u^d(x) \quad x \in \partial\overline{\Omega}_U \tag{5a}$$

$$t(x) = t^d(x) \quad x \in \partial\overline{\Omega}_F \tag{5b}$$

$$u(x_0) = 0 \tag{5c}$$

The purpose of this section is to show the following lemma:

**Lemma**

*Assumptions*

1/ u is a vector field on $\overline{\Omega} \cup \partial\overline{\Omega}$, which is $C^{0,\beta}$ ($\beta$-holderian) with $\beta > 0$

2/ u is such that L(u)=0, L being the operator of linear plane isotropic elasticity within $\overline{\Omega} \cup \partial\overline{\Omega}$

3/ u satisfies the conditions (5),

*Consequence* : u satisfies the boundary integral equation (12) in any regular point of $\partial\overline{\Omega}$ and the integral representation (13) below.

Let us consider a function u satisfying the above hypotheses and let us consider the restriction of u to $\overline{\Omega}_r$. The boundary $\partial\overline{\Omega}_r$ of $\overline{\Omega}_r$ is constituted by $\partial\overline{\Omega}$, $S_r$ and $P_r$ (Fig. 1a). Solution u satisfies a boundary integral equation (6) for any bounded open set $\overline{\Omega}_r$ [2]:

$$\int_{\partial\Omega_r} \left\{ (u_i(y) - u_i(x)) T_i^k(x,y) - t_i(y) U_i^k(x,y) \right\} ds_y = 0 \tag{6}$$

This equation (6) is valid for $x \in \overline{\Omega}_r$ and for $x \in \partial\overline{\Omega}_r$. As the elementary solution T respects the condition of null traction on $P_r$, the integration on $\partial\overline{\Omega}_r$ is reduced to the integration on $S_r$ and $\partial\overline{\Omega}$.

Let us introduce the "modified Green functions" defined below by (7,8):

$$U *_i^k (x,y) = U_i^k(x,y) - U_i^k(x_0, y) \tag{7}$$

$$T *_i^k (x,y) = (T_i^k(x,y) - T_i^k(x_0, y)) \tag{8}$$

Replacing x by $x_0$ in (6), making the difference with the original equation (6) and using (5c), leads to :

$$\int_{\partial\Omega_r} \left\{ (u_i(y) - u_i(x)) T_i^k(x,y) - u_i(y) T_i^k(x_0, y) - t_i(y) U *_i^k (x,y) \right\} ds_y = 0 \tag{9}$$

The part of the integral above on $S_r$ can be written as:

$$I_R = \int_{S_r} \{(u_i(y) - u_i(x))T_i^k(x,y) - u_i(y)T_i^k(x_0,y) - t_i(y)U_i^{*k}(x,y)\}ds_y \qquad (10)$$

One has $\int_{S_r} u_i(x)T_i^k(x,y)ds_y = -u_k(x)$ (because of the balance condition on the boundary of $\Omega^*_r$ with $x \in \Omega^*_r$, see Fig. 1b), which leads to :

$$I_R - u_k(x) = \int_{S_r} \{u_i(y)T_i^{*k}(x,y) - t_i(y)U_i^{*k}(x,y)\}ds_y \qquad (11)$$

One can check that $U_i^{*k}(x,y)$ is O(1/r) when r(y) tends to infinity, and that $T_i^{*k}$ is O(1/r$^2$). Due to Saint-Venant principle, u and t behave at infinity as the response to the resultant of the forces applied on the boundary. It means that u is O(ln(r)) and t is O(1/r). Using polar coordinates, it can be concluded that the integral given by (11) tends to 0 as r tends to infinity. By using (10), an integral equation for the relative displacement which is valid for $x \in \overline{\Omega} \cup \partial\overline{\Omega}$ is finally obtained:

$$u_k(x) + \int_{\partial\overline{\Omega}} \{(u_i(y) - u_i(x))T_i^k(x,y) - u_i(y)T_i^k(x_0,y) - t_i(y)U_i^{*k}(x,y)\}ds_y = 0 \qquad (12)$$

The boundary integral equation on $\partial\overline{\Omega}$ is the special case of (12) when $x \in \partial\overline{\Omega}$. Finally, one can write an integral representation for any point $x \in \overline{\Omega}$. Taking into account the equilibrium condition on the boundary of $\Omega$, leads to $\int_{\partial\overline{\Omega}} -u_i(x)T_i^k(x,y)ds_y = 0$ and equation (12) yields, for any point of $\overline{\Omega}$ which is not on its boundary, to:

$$u_k(x) = \int_{\partial\overline{\Omega}} \{t_i(y)U_i^{*k}(x,y) - u_i(y)T_i^{*k}(x,y)\}ds_y = 0 \qquad (13)$$

Replacing x by $x_0$ in (13) leads to $u_k(x_0) = 0$.

In conclusion, it is proved that any elastic solution in "relative displacement" satisfies particular forms of boundary integral equation and of integral representation (12, 13). It is worthwhile to mention that this integral representation ensures that the condition $u(x_0)=0$ is satisfied. However, this method does not provide a way for finding the solution of the classical formulation of the problem since the prescribed relative displacement (5a) is not known from the boundary conditions in the classical formulation (14b).

## 4. Back to the classical formulation

Instead of a supplementary condition at $x_0$, conditions on the behaviour to infinity are now considered. In fact, these conditions at infinity are naturally obtained from the usual choice of Green functions (3, 4). The integral representation corresponds indeed physically to a suitable set of forces and dipoles, whose behaviour at infinity is given by relations (3) and (4). The set (14) of conditions at the boundary and at infinity is now:

$$t(x) = t^d(x) \quad x \in \partial\overline{\Omega}_F \tag{14a}$$

$$v(x) = v^d(x) \quad x \in \partial\overline{\Omega}_U \tag{14b}$$

$$v_2(x_1, x_2 = 0) = A_2 \ln|x_1| + B_2 \mathrm{sign}(x_1) + O(1/x_1) \quad x_1 \to \pm\infty \tag{14c}$$

$$v_1(x_1 = 0, x_2) = A_1 \ln x_2 + O(1/x_2) \quad x_2 \to +\infty \tag{14d}$$

Where the constants $A_1$, $A_2$ and $B_2$ are generally not known a priori. They depend on the resultant force of the loading.

The purpose of the present section is now to prove the following theorem:

**Theorem**

*Assumptions*

1/ Let us consider a solution v of the elasticity problem $L(v)=0$ for the half plane

2/ v meets the conditions (14a) to (14d) above

3/ v is assumed to be $C^{0,\beta}$.

*Consequence:* v satisfies the integral representation (2) and the boundary integral equation (1) in any regular point of $\partial\overline{\Omega}$.

**Remark**

In other words, it is intended to obtain a boundary integral equation without assuming that $v = O(r^{-\alpha})$ and $t = O(r^{-1-\alpha})$ (with $\alpha > 0$).

Let us consider a chosen point $x_0 \in \overline{\Omega}$ ($x_0 \notin \partial\overline{\Omega}$) and the relative displacement u of any point of the domain defined as the difference to the displacement at $x_0$ (15):

$$u(x) = v(x) - v(x_0) \tag{15}$$

Then u is solution of the following auxiliary problem (16), which is a problem in relative displacements with regard to $x_0$. The traction conditions are the same as in the initial problem and the displacement conditions have been translated by $v(x_0)$.

$$t(x) = t^d(x) \quad x \in \partial\overline{\Omega}_F \tag{16a}$$

$$u(x) = v^d(x) - v(x_0) \quad x \in \partial\overline{\Omega}_U \tag{16b}$$

$$u(x_0) = 0 \tag{16c}$$

v is $C^{0,\beta}$ and u meets obviously the same property. The lemma of section 3 indicates that u meets the integral equation (12) and the integral representation (13).

Equation (12), valid for $x \in \overline{\Omega} \cup \partial\overline{\Omega}$, is now considered. Replacing u by (15) in (12) leads to (17), noting that the integrals can be split into two parts because the right-hand side of (17) has no singularities ($x_0 \notin \partial\overline{\Omega}$):

$$\begin{aligned} v_k(x) + \int_{\partial\Omega} \{(v_i(y) - v_i(x))T_i^k(x,y) - t_i(y)U_i^k(x,y)\}ds_y = \\ v_k(x_0) + \int_{\partial\Omega} (v_i(y) - v_i(x_0))T_i^k(x_0,y) - t_i(y)U_i^k(x_0,y)ds_y \end{aligned} \tag{17}$$

The right-hand side of (17) does not depend on x. It can be shown that it is equal to zero as follows. Assuming that $x \in \overline{\Omega}$, (17) can be rewritten:

$$v_k(x) + \int_{\partial\overline{\Omega}} v_i(y) T_i^k(x,y) - t_i(y) U_i^k(x,y) ds_y$$
$$= v_k(x_0) + \int_{\partial\overline{\Omega}} (v_i(y) - v_i(x_0)) T_i^k(x_0,y) - t_i(y) U_i^k(x_0,y) ds_y \tag{18}$$

In the left-hand term, for k=1, due to the classical choice of the Green function, the property $U_i^1((x_1 = 0, x_2), y) = A_i \ln(x_2) + O(1/x_2)$ is met when $x_2$ tends to infinity. The integral $\int_{\partial\overline{\Omega}} v_i(y) T_i^1(x,y) ds_y$ tends to zero when r tends to infinity and due to (14d), it can be concluded that the left-hand side can be written as $B\ln(x_2)+o(1)$ when $x_1=0$ and $x_2$ tends to infinity. As the right-hand side is constant, we conclude that B is zero and that the right-hand side is zero for k=1 (4.6).

$$v_k(x_0) + \int_{\partial\overline{\Omega}} (v_i(y) - v_i(x_0)) T_i^k(x_0,y) - t_i(y) U_i^k(x_0,y) ds_y = 0 \tag{19}$$

A similar proof can be used for k=2, using (3) and (14c). From (17) and (19), it can be deduced that u is solution to the boundary integral equation (1) for $x \in \partial\overline{\Omega}$ and of the integral representation (2) for $x \in \overline{\Omega}$. These results do not depend on the choice of $x_0$.

The integral representation (2) makes it possible to check that v satisfies the condition to infinity (14c) and (14d), due to the fact that the Green functions satisfy conditions (3) and (4).

## 5. Conclusions

In a first step an exterior elastostatic problem in a half plane was studied by replacing the usually considered conditions at infinity by a condition of zero displacement at a chosen point, not located on the boundary. If the solution is assumed regular enough, it meets specific forms of a boundary integral equation and of an integral representation without any artificial restrictive hypotheses on the behaviour to infinity.

If one assumes in a second step that u satisfies specific conditions to infinity, and that u is $C^{0,\beta}$, it has been proved that u satisfies the usual forms of boundary integral equation and integral representation written on the finite boundary. Such a result was proved up to now only under largely too restrictive decreasing conditions at infinity. It can be mentioned that similar results are obtained along similar lines for the full plane problem.